# How Many Hydrated NEOs Are There?

## Andrew S. Rivkin[1] and F. E. DeMeo[2]

1. Johns Hopkins University Applied Physics Laboratory

    11101 Johns Hopkins Rd

    Laurel MD, 20723 USA

    443-778-2811

    andy.rivkin@jhuapl.edu

2. Massachusetts Institute of Technology

## 1. Abstract

Hydrated minerals are tracers of early solar system history, and have been proposed as a possible focus for economic activity in space. Near-Earth objects (NEOs) are important to both of these, especially the most accessible members of that community. Because there are very few identified hydrated NEOs, we use the Ch spectral class of asteroids as a proxy for hydrated asteroids, and use published work about NEO delivery, main-belt taxonomic distributions, NEO taxonomic distributions, and observed orbital distributions to estimate the number of hydrated asteroids with different threshold sizes and at different levels of accessibility. We expect $53 \pm 27$ Ch asteroids to be present in the known population of NEOs > 1 km diameter, and using two different approaches to estimate accessibility we expect $17 \pm 9$ of them to be more accessible on a round trip than the surface of the Moon. If there is no need to define a minimum size, we expect $700 \pm 350$ hydrated objects that meet that accessibility criterion. While there are few unknown NEOs larger than 1 km, the population of smaller NEOs yet to be discovered could also be expected to contain proportionally-many hydrated objects. Finally, we estimate that hydrated NEOs are unlikely to bring enough water to account for the ice found at the lunar poles, though it is possible that asteroid-delivered hydrated minerals could be found near their impact sites across the lunar surface.

## 2. Introduction

A subset of meteorites has abundant minerals with OH and/or $H_2O$ (hereafter simply called "hydrated minerals" for convenience). These are of great interest to the planetary science community for several reasons: they are a tracer of physical and chemical conditions early in solar system history (Bell et al., 1989), and they demonstrate that asteroid impacts could have brought a significant amount of water to the young terrestrial planets, (Morbidelli et al. 2012). They also are examples of relatively weak material that has reached the Earth's surface and thus provide a loose constraint on how strong material must be in order to survive entry. The meteorites we collect on Earth only give us part of the overall story. The added details available from study of returned samples from the Hayabusa2 and OSIRIS-REx missions will deepen our understanding greatly. However, to fully satisfy the community interest, we also need the added context of knowing the distribution of hydrated minerals in the present-day asteroid population. This is most commonly and efficiently done by linking meteorites to asteroids via telescopic observations.

Beyond the planetary science interest in asteroidal hydrated minerals, there is also interest in them from the exploration community (Graps et al. 2016, Sercel et al. 2018). The last decade has seen the establishment of private companies interested in prospecting asteroids and several countries are passing laws that encourage (or at least allow for) mining them (Foster 2016, Davis and Sundahl 2017, Kfir and Perry 2017). Hydrated minerals are candidate materials for establishing a space-based mining economy (Metzger 2017), with the application of using them in space as a means of refueling communications satellites since it is energetically easier to bring material to geosynchronous orbit from many near-Earth asteroids than from the surface of the Earth. Space agencies have also taken interest in resource extraction and utilization, and in-space resource utilization (ISRU) experiments are ongoing on Earth (Gertsch et al. 2017, Linne et al. 2017, Dreyer et al. 2018) and were proposed for the Asteroid Redirect Mission (Mazanek et al. 2014, Elliott et al. 2015). Remote sensing of asteroid compositions using astronomical techniques is the most efficient way to identify candidate objects of the most interest to the ISRU/mining communities and to understand how common those objects are.

There is evidence for aqueous alteration in a large fraction of the meteorite classes (Brearly 2006). However, much of this evidence is coupled with evidence of subsequent metamorphism and destruction of hydrated minerals (Grossman et al. 2000). Brearly (2006) notes that in many of the meteorite groups alteration did not result in widespread formation of hydrous phases. To generalize, hydrated minerals that can be convincingly detected via remote sensing are largely confined to the CM and CI carbonaceous chondrite groups. These groups have water concentrations of ~10% by mass, though individual samples have water concentrations ranging from ~2-25% (Jarosewich 1990, Alexander et al. 2013, King et al. 2015).

Here we attempt to estimate the answer to a simple question: How many hydrated asteroids should we expect to find in the NEO population? A promising approach, used below as well as by previous studies, is akin to the famous Drake Equation approach for estimating the number of civilizations in the galaxy (Drake 1962): decompose the final number into the product of several factors, each of which in turn can be calculated or more easily estimated.

Sanchez and McInnes (2011) made an estimate of the liters of water available for mining in the NEO population, but were focused on dynamical and engineering topics and did not consider the number of objects this water would be contained in. Elvis (2014) also addressed this question from a somewhat more engineering and economics-based angle than we use, quantifying the number of "ore-bearing asteroids" ($N_{ore}$) as:

$$N_{ore} = P_{ore} \times N(>M_{min})$$

Where $P_{ore}$ is the probability the object is ore-bearing and $M_{min}$ is the "minimum profitable mass".

$P_{ore}$ in turn is defined as the product of other probabilities $P_{type}$, $P_{rich}$, $P_{acc}$, and $P_{eng}$, which are the probabilities that an asteroid is of the correct composition (see next section), that it is sufficiently rich in the resource in general, that it is in an accessible orbit, and that ore can be profitably extracted. Elvis estimated the fraction of water-rich NEOs as 2.5% based on the estimates of NEO taxonomic type distribution from Stuart and Binzel (2004) and the water concentrations in carbonaceous chondrites found by Jarosewich (1990). His estimates for $N_{ore}$ are sensitively dependent upon the size and delta-v cutoffs used (see Section 7 for a fuller discussion of delta-v, which is a measure of the change in speed needed to send a spacecraft from one orbit to another specified orbit), but he estimates roughly 18 objects with absolute magnitude (H) < 22 (roughly 167-264 m diameter for albedos 4-10%, relevant to carbonaceous chondrites. Unless otherwise noted, we will use an albedo of 6% for Ch asteroids in this paper).

A very recent paper by Jedicke et al. (2018) also addresses the question of the number of asteroids that can be accessed for water extraction, though they focus on very small objects (5-10 m in diameter) with very small values of delta-v that can be brought back to a distant retrograde lunar orbit. They find "thousands" of objects that meet their criteria and are water-rich. We find the fraction of hydrated NEOs to be consistent with their result, but it is difficult to make a direct comparison of numbers because of the large uncertainties in population size at such small diameters and because it is not straightforward to compare the simple dynamics used in this work to the more rigorous dynamical investigation they undertake.

Below we reconsider the question from the astronomical point of view, without addressing any engineering or economic questions and only lightly touching upon dynamical ones. However, we will consider a range of astronomical and meteoritical observations and measurements to provide a sense of not only the likely order of magnitude of objects of interest, but the uncertainty on that number and how the uncertainty can be reduced. We focus on 100-m scale and 1-km scale objects, and comparisons to the Moon and to asteroid (101955) Bennu as accessibility criteria.

In Section 3 we motivate using the Ch class of asteroids as representative of hydrated objects. In Section 4 we address the "supply side" of hydrated NEOs by looking at the fraction of Ch asteroids in the main belt and current models for NEO delivery. In Section 5 we consider what is seen in the NEO population itself. The final sections consider how the numbers from Sections 4 and 5 differ and possible reasons why, the number of accessible hydrated NEOs and the amount of water they contain, implications of the hydrated NEOs to supply of ice to lunar permanently shadowed regions, and future work.

## 3. Hydrated Minerals and Ch Asteroids

Currently there are two primary ways of spectroscopically detecting hydrated minerals on asteroid surfaces: via an absorption band in the 3-$\mu$m region and via an absorption band near 0.7 $\mu$m (Figure 1). The former is due to hydroxyl and water, and so is a direct measurement of those materials in an asteroidal regolith. Hundreds of observations of main-belt asteroids have been made at these wavelengths, and hydrated minerals have been detected on NEOs from their 3-$\mu$m spectra as well (Volquardsen et al. 2007, Rivkin et al. 2013, Rivkin et al. 2018). Unfortunately, it is also a somewhat difficult wavelength region to work in for several reasons. First, water in the Earth's atmosphere effectively precludes ground-based measurements in the ~2.5-2.85 $\mu$m region even in the best observing sites, while the band minimum for OH-bearing minerals is typically found in that wavelength range. Second, while L-band filters in the 3-4 $\mu$m wavelength region are not completely standardized, they typically exclude wavelengths shortward of ~3.2 $\mu$m (for instance, Tokunaga et al. 2002), which is where the deepest absorptions in hydrated minerals are found (Figure 1). Compounding this, L-band filters are often used as blocking filters in astronomical spectrographs, so many instruments are incapable of covering the wavelengths of interest. The combination of limited suitable observing sites and appropriate instrumentation has limited 3-$\mu$m asteroid observations to a small number of telescopes, and useful measurements of 3-$\mu$m band depth have been effectively limited to objects with V magnitude < 15, a brightness reached by few NEOs. Unfortunately, an additional issue often manifests for those objects: NEOs with V < 15 must typically be at solar distances ~1 AU, which results in prodigious thermal emission, which in turn makes models more important for removing thermal flux and allowing a reflectance-only spectrum to be analyzed. At sufficiently-high temperatures, enhanced thermal flux due to increased emissivity at wavelengths with reduced reflectance can lead to absorption bands being removed entirely (Clark 1979). Future missions like the James Webb Space Telescope will be able to observe much fainter objects in the 3-$\mu$m region, but observing time will likely be in short supply (Rivkin et al. 2016).

The second indicator of hydrated minerals in asteroids, and the one we use here, is at 0.7 $\mu$m. This absorption is due to a charge transfer in oxidized iron, and while it is not diagnostic of hydrated minerals *per se* it has only been seen on asteroids and in meteorites in conjunction with a 3-$\mu$m band (Vilas and Gaffey 1989, Rivkin et al. 2015). The presence of the 0.7-$\mu$m band is the defining characteristic of the Ch asteroid class in the Bus-DeMeo classification (DeMeo et al. 2009), and one of two defining characteristics of the Cgh asteroid class. Thus, every Ch and Cgh asteroid can be safely assumed to have hydrated minerals. In addition to this, measurements of Ch asteroids in the 3-$\mu$m region by Rivkin et al. (2015) show that Ch/Cgh asteroids all have 3-$\mu$m band shapes that are like those we see in meteorites, which are caused by phyllosilicates and other minerals that are stable at 1 AU on airless body surfaces. Note that the Tholen (1984) taxonomy is still in some use in parallel with the Bus-DeMeo taxonomy, and does not use the 0.7-$\mu$m band or have a Ch class. We will specify when this taxonomy is being used in this paper.

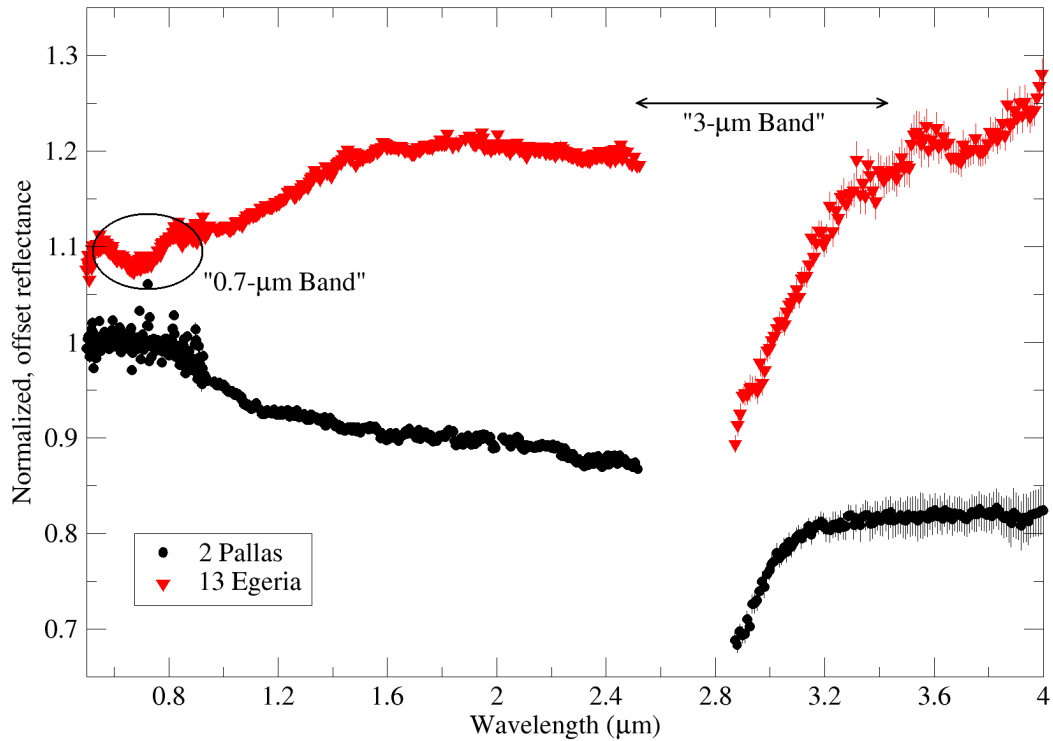

**Figure 1:** Spectra of two large, low-albedo asteroids show the absorption bands associated with hydrated minerals. Water vapor in the Earth's atmosphere makes the 2.5-2.8 μm region effectively opaque, and data in that region is omitted. The "3-μm band" has its minimum within this opaque region, but is sufficiently broad that it is measureable longward of the opaque region. The 0.7-μm band that we use in this work as a signifier of hydrated minerals is circled in the spectrum of 13 Egeria (classified as a Ch-type asteroid) while it is absent in the spectrum of the B-type asteroid 2 Pallas. The 0.5—2.5 μm portions of these spectra are from DeMeo et al. (2009), the 2—4 μm portions for Egeria and Pallas are from Rivkin et al. (2015) and Rivkin et al. (2011), respectively.

This is important because this is not thought to be the case for all asteroids with 3-$\mu$m absorption bands. In particular, several low-albedo outer-belt asteroids in the B class

(including 24 Themis) and C class have been identified with 3-$\mu$m band shapes that are best fit by ice frost (Rivkin and Emery 2010, Campins et al. 2010, Takir and Emery 2012). Carry et al. (2016) found 54 near-Earth asteroids observed by the Sloan Digital Sky Survey (SDSS) with colors consistent with a C-class taxonomy and 4 with B-class colors, and the Binzel et al. (2004) spectral survey found 13 C-class and 5 B-class objects, but no objects with Themis-like 3-$\mu$m band shapes have been identified among the NEOs at this writing. Conditions on NEO surfaces are much too hot to allow the presence of near-surface ice, and while it is theoretically possible that recent arrivals to NEO space with sufficiently-insulating and deep regolith could maintain ice very deep in their interiors (Schorghofer and Hsieh 2018), there is as yet no obvious way to identify such objects. Focusing on Ch/Cgh asteroids (hereafter collectively termed "Ch asteroids" in this paper for narrative convenience) allows would-be prospectors to reduce uncertainty about what they'll find on an asteroid surface and design extraction techniques and economic cases around a reasonably narrow set of expectations, and also allows us to use these objects to trace aqueous alteration in the asteroid population.

For completeness, we note that while the 0.7- and 3-$\mu$m regions are the primary ways of identifying hydrated asteroids, there is also ongoing work using mid-infrared (8-25 $\mu$m) measurements to characterize hydrated minerals in meteorites and asteroids (Beck et al. 2014, McAdam et al. 2015, King et al. 2015). While this has great promise, the current challenges facing a mid-IR asteroid spectral survey are similar to those that face 3-$\mu$m observations.

The CM meteorites are thought to be the best analog to the Ch asteroids. Several lines of evidence link the CMs to the Ch asteroids, including the fact that the vast majority of CM meteorite spectra feature a 0.7-$\mu$m band, while it is rare or absent in the spectra of members of other meteorite groups. Indeed, other than the CM meteorites, few meteorite groups have hydrated minerals at all, though the CI chondrites have a range of water concentrations that overlap with the CM group (Alexander et al. 2013, King et al. 2015). Band depths and shapes in the 3-$\mu$m region are consistent between the Ch and CM meteorites, and the overall albedo is also similar, and several authors have concluded that these groups are effectively equivalent if not perfectly identical sets (Vilas 1994, Carvano et al. 2003, Fornasier et al. 2014, Rivkin et al. 2015, Vernazza et al. 2016). Nevertheless, we note before proceeding further that the number of Ch NEOs will be a lower limit of objects of interest because there are hydrated meteorites that do not have the 0.7-$\mu$m band (CR Renazzo and CI Orguiel, for instance) and hydrated NEOs that are not Ch asteroids ((175706) 1996 FG3, for instance, has a measured 3-$\mu$m absorption band but no 0.7-um band). Similarly, near-Earth objects derived from the cometary population may have interior ice (though only under very limited circumstances: Schorghofer and Hsieh 2018), but are not expected to have phyllosilicates and thus would not be classified as Ch-class objects.

## 4. Predicted Ch Asteroid Supply to NEO space

While we are unaware of any estimates in the literature of what fraction of asteroids are Ch asteroids, nor what fraction of NEOs should be Ch asteroids, the estimate is fairly simple to make and the components are present in the literature. Rivkin (2012) used two different statistical techniques to estimate that roughly 30% ± 5% of 3724 main belt C-complex asteroids in the Sloan Digital Sky Survey (SDSS) had a 0.7-$\mu$m absorption. This was broken down further, with the Ch-like fraction of C-complex objects in the inner, middle, and outer belts estimated as 29% ± 4%, 37%± 4%, and 28% ± 5%, respectively. Different approaches in Rivkin (2012) are consistent with either a decrease in Ch fraction with decreasing size or no change with size. In either case the Ch fraction in the smallest size bin investigated (H 16.0-18.0, or diameter ~1.5-3.4 km for albedo of 6%) is consistent with the whole-belt, all-size value of 30% ± 5%.

Fornasier et al. (2014) found 50% of the C-class objects (using the Tholen taxonomy) in their main-belt sample to have 0.7-$\mu$m bands, but found related classes in that taxonomy to have varying percentages of hydrated asteroids: 7.7%, 9.0%, and 100% for the F, B, and G classes, respectively. The percentage of C, F, B, G asteroids (taken as a single set) with 0.7-$\mu$m bands in the Fornasier et al. sample is 45%. It is not clear whether there is a bias such that one class was over- or undersampled relative to the others, and a change in the B:C ratio from the reported 92:454 would strongly affect the overall estimated hydrated asteroid fraction. Fornasier et al. also reported hydrated fractions of their sample vs. semi-major axis, though not with the same bounds as Rivkin (2012). An apples-to-apples comparison is not straightforward, and the objects reported in Fornasier et al. include not only C-complex objects but 22 objects in the Tholen P class, but those are only a small addition to the overall sample. If we re-cast the Fornasier et al. sample into the inner, middle, and outer belt bounds from Rivkin (2012), we estimate 49%, 49%, and 36% hydrated objects within the C complex, respectively.

Fornasier et al. argued that the Rivkin numbers were likely an underestimate due to non-optimal placement of filters in the SDSS survey vs. a goal of measuring or detecting the 0.7-$\mu$m band. Nevertheless, for our purposes we will take values midway between the Rivkin (2012) and Fornasier et al. (2014) values in an attempt to represent some of the uncertainties involved in our estimation. We adopt a hydrated C-complex asteroid fraction of 39% ± 10%, 43% ± 6%, and 32% ± 6% for inner, middle, and outer belt.

These numbers can be convolved with models of the supply of NEOs from various small body reservoirs (Bottke et al. 2002). Bottke et al. estimate 61% of the NEO population is derived from the inner main belt, with the central and outer main belt providing 24% and 8%, respectively. The remaining 6% is from the Jupiter Family Comet region, for which we set the Ch fraction to 0. Uncertainties are not provided by Bottke et al. for these estimates, but they can be calculated using other information in the paper as 9%, 5%, and 1% for the inner, middle, and outer belt, respectively. The final necessary piece for this estimate is the fraction of asteroids belonging to the C complex in each reservoir. Bus and Binzel (2002) provided a debiased estimate of the fraction of each spectral complex vs. semi-major axis in the main belt for objects > 20 km. The averages for the C complex in the inner, central, and outer belt are 37%, 47%, and 52%, respectively. The implied Ch percentage among asteroids in those regions are 14% ± 4%, 20% ± 3%, and 17% ± 4%. For the belt as a whole, the Ch percentage is 17% and the C-complex percentage is 48% when weighting by the number of objects > 5 km in each region of the belt from DeMeo and Carry (2014). The model of DeMeo and Carry itself estimates C-complex asteroids to make up 49%, 53%, and 67% of inner, middle, and outer-belt asteroids by number > 5 km diameter, and 63% of the belt as a whole.

It is relatively straightforward to combine the numbers mentioned above into an estimate for the expected fraction of Ch asteroids among the NEO population as a whole:

*NEO Ch fraction = Σ [(f$_{NEO,pl}$)×(f$_{Ccomp,pl}$) ×(f$_{Ch,pl}$)]*

Where f$_{NEO}$, f$_{Ccomp}$ and f$_{Ch}$ are the fractions of NEOs derived from a particular part of the main belt, the fraction of asteroids in that part of belt that are C complex, and the fraction of C complex asteroids in that part of the belt that are Ch class, and this product is summed for all parts of the belt.

With the numbers outlined above, we calculate:
*Ch fraction = (0.61)(0.37)(0.39) + (0.24)(0.47)(0.42) + (0.08)(0.52)(0.32) = 0.15*

By this estimate, and including the uncertainties above, 15% ± 7% of all NEOs should be Ch asteroids using the Bus and Binzel main-belt compositional distribution. The C-complex should be ~38% of NEOs via this same calculation. Using the DeMeo and Carry (2014) distribution, the NEOs should be 19% Ch and 48% C-complex. Averaging these numbers and accounting for uncertainty due to the difference between the Bus and Binzel (2002) and DeMeo and Carry (2014) estimates of the C-complex fraction in the main belt, the estimate based on main-belt measurements and delivery models, with no input from observations of the NEO population itself, would lead us to expect 17% ± 3% Ch-class NEOs and 43% ± 6% C-complex objects. We note below that there are several implicit and explicit assumptions in these estimates.

## 5. Observed Ch Asteroids in NEO Space

Given the estimate from the previous section, based on NEO delivery models and the taxonomic makeup of the main asteroid belt, we might expect hundreds of C-complex asteroids and well over 100 specifically Ch-class asteroids > 1 km in diameter in the NEO population. However, the number of known Ch asteroids in NEO space is very small. In the nearly 2300 asteroids in the European Asteroid Research Node database (http://earn.dlr.de/nea/table1_new.html, 1 Feb 2018 update), 740 of which have taxonomic types assigned at this writing, only 3 are identified as Ch asteroids: (285263) 1998 QE2, 2002 DH2, and 2012 EG5. This very small fraction is highly affected by biases, of course, among them the popularity of 0.8-2.5 $\mu$m observations of NEOs, which cannot be used to identify Ch asteroids, eclipsing the popularity of 0.5-1.0 $\mu$m observations, which can. The most recent large survey of NEOs in the visible-near IR (Binzel et al. 2004) reported 1 Ch asteroid out of 23 C-complex NEOs, which is again much smaller than 17% but may be affected by the statistics of small numbers. Surveys of the NEO region specifically find C-complex fractions that are smaller than the 43% estimated above. Stuart and Binzel (2004) report a C-complex fraction of 9.8% ± 3.3% in the NEO population, while Carry et al. (2016) used SDSS data to find 23% of 230 NEOs were C-complex asteroids. These two numbers suggest a Ch fraction of 6% ± 3% in the

NEO population, including the uncertainty due to the discrepancy between the Stuart and Binzel and Carry et al. estimates.

Another possible data source is the NEOWISE survey, though the contents are ambiguous on this topic. Using the bounds for taxonomic classes published in Mainzer et al. (2012) would suggest that 12% of the 778 different NEOs in their full sample (Mainzer et al. 2016) would be classified as C or B, and 17% of the 275 different NEOs > 1 km. However, this dataset has not been debiased. The same dataset shows the low albedo ($p_v < 0.1$) percentages to be 36% and 43% of the total and > 1 km populations, respectively, suggesting the C complex constitutes ~33-40% of low-albedo NEOs. Using the same approach on the main-belt dataset gives unexpected results, however. While the set of measurements of objects > 5 km diameter shows 63% to be low albedo, only 9% are classified as C or B. Given the C-complex fraction of low-albedo NEOs, we would have expected ~20-25% C or B asteroids in the NEOWISE main-belt asteroid dataset. This data set includes duplicate observations of objects, but with well over 50,000 measurements this seems unlikely to cause this discrepancy. Solving this particular problem is beyond the scope of this paper, and we do not further consider this question, but we do note that the NEOWISE data has potential for greatly increasing our knowledge of the taxonomic distribution of main-belt asteroids.

When we look to the meteorites, both the implied Ch and C-complex fractions in the NEO population are smaller still. The Meteoritical Bulletin database (https://www.lpi.usra.edu/meteor/metbull.php, 9 Feb 2018 update) reports 16 CM falls and 45 carbonaceous chondrite falls out of 1153 total falls of non-Mars meteorites (there are no recorded falls of lunar meteorites). The CMs making up ~1/3 of all carbonaceous chondrite falls is consistent with the estimates above from astronomical measurements of asteroids. However, taken at face value the fall statistics imply that less than 2% of NEOs are CM and only 4% of NEOs are carbonaceous. The extent to which the meteorite fall fraction represents the fraction of Ch asteroids in space is discussed further in the next section.

Table 1 gathers and summarizes the various input parameters and the resulting estimates that we calculate in this section.

## 6. Is there a C Asteroid Problem?

It is striking to see that there is nearly an order of magnitude discrepancy between the Ch fraction of main-belt asteroids estimated in the sections above and the fraction of meteorite falls associated with the best analog to the Ch asteroids, the CM meteorites. Perhaps the most unexpected discrepancy is the factor of 2-3 between the fraction of C-complex asteroids seen in the NEO population and that expected given observations and our understanding of NEO delivery. Indeed, because our estimates of the Ch fraction are as a fixed percentage of the C-complex population, the discrepancy in the observed vs. expected C asteroid fraction is the main driver of our overall uncertainty.

We can imagine several possible reasons for these discrepancies, and possible processes that could filter out Ch and/or C-complex objects at several points along the delivery path. These processes are of varying applicability and interest to the scientific and ISRU communities and include differential orbital evolution, differential lifetimes for hydrated and anhydrous objects, unrecognized (or recognized) bias in populations, and preferential removal of weak material. Figure 2 schematically shows some of the processes that could have affected the C-complex and Ch-class/CM meteorite percentages from the main belt to the ground.

Thoroughly discussing these factors is out of the scope of this paper, but we touch on them here:

### 6.1. Bias in Delivery?

The calculation above assumes that there is no collisional or dynamical preference for one composition over another, and that the resonances that move material into near-Earth orbits are fed objects via Yarkovsky drift in amounts representing the average main-belt composition. Furthermore, there is an implicit assumption that the size-frequency distributions for different asteroid classes are the same—that the relative fraction of C-complex asteroids at 100-m sizes is the same as that fraction at 10-km sizes, say. It has long been noted that the median composition of a cosmic dust particle is quite different from that of a median meteorite fall (Bell 1991) and Vernazza et al. (2017) proposed a link between cosmic dust and C-complex asteroids. Housen et al. (2018) recently looked at impacts into porous targets, concluding that ejecta is suppressed for high-porosity objects above ~50 km diameter. If main-belt C-complex objects have preferentially higher porosity than S-complex objects (Consolmagno et al. 2008), they might be expected to create less ejecta on average, and provide less material for transport to NEO space than otherwise expected.

### 6.2. Observational Bias?

The low albedos of C-complex asteroids make them fainter than S-complex asteroids of the same size, and thus more difficult to detect. This is a well-known effect, but it can be difficult to correct in data (Rabinowitz 1994, Michel et al. 2000, Jedicke et al. 2002, Stuart and Binzel 2004). The Stuart and Binzel estimate of C-complex NEO abundance depends in part on the X-complex abundance, which in turn is dependent upon some subjective factors (the boundary between C- and X-complexes) and some uncertain ones (the relative number of X-complex asteroids that are low albedo vs. those that are high albedo). The task of debiasing the NEOWISE dataset is still underway, though its detections should be less sensitive to albedo biases than visible-wavelength surveys (Mainzer et al. 2015). As noted above, NEOWISE measured ~40% of NEOs to be low albedo, which is consistent with the Stuart and Binzel estimate of ~44% ± 15% low albedo NEOs (where they assumed the X complex to be split 55%-45% between high-albedo and low-albedo objects).

### 6.3. Preferential Disruption?

There is mounting evidence for non-collisional processes that can remove NEOs from the population. Marchi et al. (2009) recognized that many NEO orbits go through periods of low perihelia, leading to increased heating. The C-complex asteroid 3200 Phaethon is in such a low-perihelion orbit, and is thought to be shedding material via dehydration and associated stresses (Jewitt et al. 2013). Granvik et al (2016) find that there are fewer small, low-perihelion asteroids than models would predict, suggesting that processes similar to what is seen on Phaethon are removing objects, preferentially those objects that are hydrated and/or low-albedo. This would also lead to the presence of fewer C-complex objects in the NEO population than one might expect based on the main-belt population.

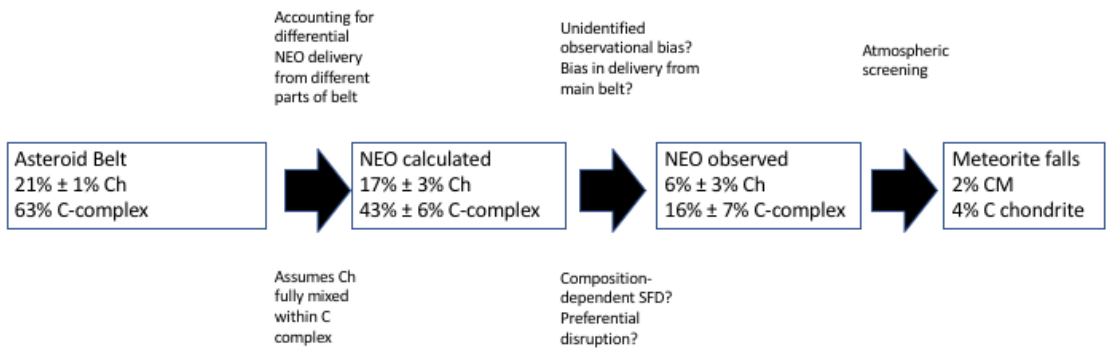

**Figure 2:** Schematic representation of the path asteroids take to near-Earth space and ultimately to become meteorites, with the observed or calculated Ch fraction and factors that may change or lead us to misestimate that fraction.

### 6.4. Atmospheric Screening

The idea that the Earth's atmosphere filters out weaker material from reaching the surface is not controversial, and for a given size stronger meteorites like irons are more likely to reach the surface intact and be recorded as a fall than weaker material (Britt et al. 2016). This can be seen by looking at the interplanetary dust particle (IDP) collection. IDPs < 100 $\mu$m in size and collected at altitude have characteristics that differ from the larger particles (and meteorites) that reach the ground (Messenger et al. 2013). One of the three major compositional-textural groups of IDPs (the chondritic, smooth and compact particles: CS) are composed mostly of hydrated minerals (Schramm et al. 1989). The CS IDPs are broadly similar to the CM and CI chondrites and are thought to have asteroidal origins, but it is not clear if they are from literally the same parent bodies as the larger CM and CI meteorites. One of the other major IDP groups, the chondritic-porous (CP) group, is seen to be extremely fragile and often do not survive collection or sample preparation intact (Messenger et al. 2013).

The relative fluxes of the CS, CP, and the third major group (single mineral grains) are uncertain, but it is evident that the fraction of IDPs in the CS group is much higher than the fraction of CM or CI falls, and conversely IDPs with geochemistry consistent with the ordinary chondrite meteorites are much less abundant than they are in the fall statistics (Brownlee et al. 1997). Experiments by Tomeoka et al. (2003) and Flynn et al. (2009) showed that CM targets fracture more readily and create much greater amount of mg-to-$\mu$g-scale fragments upon impact than anhydrous meteorites. Disentangling the host of processes that differentially act upon hydrated vs. anhydrous meteoroids is well beyond the scope of this paper, but it seems clear that relying on the relative fraction of cm-scale or larger CM meteorites seen to fall may cause us to vastly underestimate the number of CM parent bodies in space.

## 7. Numbers and Accessibility

Despite the complicated, intertwined processes and uncertainties discussed above, we can press forward keeping these uncertainties in mind. If we disregard the meteorite fall statistics as unrepresentative of what we might find in space, and accept that an estimate based directly on observations of NEOs is preferred to an estimate based on theoretical models of NEO delivery and measurements of main-belt asteroids, our best estimate for Ch fraction among NEOs is 6 ± 3% (Table 1). We note that if unidentified observational bias causes the discrepancy (Figure 2), the true fraction of Ch-class NEOs could be closer to the 17% estimate than the 6% estimate. Nevertheless, for the rest of this section we adopt the estimated 6 ± 3% Ch fraction and recognize the numbers are lower limits. With 886 known NEOs larger than ~1 km in diameter (Chamberlin, 2018), a size range that is thought to be ~95% discovered for NEOs, this estimated fraction suggests 53 ± 27 Ch asteroids larger than that size (rounding to the nearest "whole" asteroid), with correspondingly more at smaller sizes.

We can also look at accessibility. This is, surprisingly, not as straightforward as one might suspect. The energy required for a spacecraft to change orbits is typically reported in terms of "delta-v", the change in speed (not technically a change in velocity, despite its name) required to get into the desired orbit from the current orbit. The minimum delta-v necessary to reach a specific target is dependent upon factors such as the availability of planetary encounters, which can significantly reduce the amount of propellant that needs to be carried. While it is relatively easy to calculate the delta-v needed if there are no constraints on launch timing or mission duration, it is much more time consuming to calculate actual trajectories with realistic constraints. Below, we consider both a less-rigorous, constraint-free case and a more rigorous but restricted case extrapolated from work done in support of mission design to NEOs. Both cases provide numbers well within a factor of 2 of one another, and are consistent with one another within uncertainties. Both of these approaches assume an impulsive spacecraft; low-thrust missions would likely increase the number of accessible targets further.

### 7.1. Shoemaker and Helin-based estimates

Benner, following the work of Shoemaker and Helin (1978), provides a list of the delta-vs necessary to rendezvous with over 18,000 NEOs (https://echo.jpl.nasa.gov/~lance/delta_v.rendezvous.html, 21 June 2018 update). These calculations are not designed to calculate two-way trips, and are not strictly applicable. However, as will be shown below they do not give results that are terribly different than a more rigorous calculation provides, and the Benner work has the benefit of including data for every known near-Earth asteroid.

One possible comparison metric, of interest to the science and exploration communities, is to visits to the lunar surface. A round trip from low-Earth orbit (LEO) to the lunar surface requires ~11.4 km/s delta-v, which is a possible threshold for an object to be "more accessible than the Moon" for purposes of using extraterrestrial material in ISRU. The delta-v for a mission from low Earth orbit to the lunar surface and a direct return is roughly 9 km/s (Barbee, personal communication), with the additional delta-v required to enter LEO.

The simplest approach to the problem might be to double the one-way rendezvous delta-v to obtain the round-trip delta-v, but this would vastly overestimate what is needed. We compared the one-way delta-vs compiled by Benner to the more precise round-trip delta-vs calculated by the NASA Ames Trajectory Browser (Section 7.2) for seven objects: (162173) Ryugu, (25143) Itokawa, (101955) Bennu, (175706) 1996 FG3, (285263) 1998 QE2, (433) Eros, and (2101) Adonis. Note that the Trajectory Browser will not return results with round-trip delta-v > 10 km/s. Because we are interested in low-delta-v trajectories, this should not affect our results. The difference between the two approaches was 0.02 ± 0.42 km/s, suggesting that the one-way numbers are a sufficient representation of accessibility for our purposes. In order to account for any systematic differences that would require a more in-depth comparison we reduce the delta-v limit for objects in this section to 8.0 km/s rather than 9 km/s, in recognition that the largest differences between the two datasets were ~700 m/s and attempting to be overly conservative by allowing for a systematic offset of 1000 m/s (though, again, no such offset is seen).

Most of the asteroids we consider here do not have known sizes, so we follow typical practice and use a cutoff in H as a proxy for size. However, the appropriate choice for this cutoff is not obvious: typically H ~ 17.5 is used to represent a 1-km diameter object, which corresponds to an albedo of ~0.17. This is much higher than typical C-complex asteroid albedos and will effectively filter out C-complex asteroids < 1.6 km rather than 1 km. Using an albedo of 0.1 leads to an H value of 18.2 for a 1-km objects, but will still exclude a large fraction of C-complex asteroids of lower albedo and will additionally include smaller high-albedo asteroids. We use the albedo of 0.17 in this section, with the recognition that the resulting estimates are again lower bounds for a population for which we have consistently been using lower bounds throughout this paper. There are 270 NEOs with H < 17.5 with delta-v < 8.0 km/s according to Benner, with (67367) 2000 LY27 the most accessible large NEO. Applying the estimated Ch percentages from above to this number (and rounding to nearest whole number) gives an estimate of 16 ± 8 "accessible" Ch asteroids > 1 km assuming there is no bias for or against particular compositions in particular orbits.

We can also look to smaller objects. While it is thought that practically all of the 1-km and larger NEOs have been identified, the known fraction of smaller objects is much less. There are 4487 *known* objects with delta-v < 8.0 and H < 22.5, corresponding to ~100 m diameter for the average NEO albedo. Making the same assumptions as above, this suggests 269 ± 135 Ch-class objects in this size range have been discovered but not characterized.

Estimating the number of *undiscovered, accessible* asteroids in this way can also be done, but is less certain still. Roughly 8000 NEOs > 100 m are known. The most recent NEO models (Harris and D'Abramo. 2015) suggest ~38,000 such objects > 100 m size should exist (see further discussion in Section 7.2). If the undiscovered objects have a similar distribution of delta-vs as the discovered ones, we might expect there are ~5 times as many 100-m objects yet to be discovered. However, it is possible that the discovery rate is biased toward discovering more accessible objects since they have more Earth-like orbits.

If we remove any size constraint, there are over 11,785 known NEOs with delta-v < 8.0 km/s. If size is no object, so to speak, we estimate roughly 700 ± 350 Ch-class asteroids in the known asteroid population that are more accessible than the surface of the Moon. Finally, we recognize that the initial focus for prospective prospectors and mission planners will be on objects with delta-v much lower than 8.0 km/s, even if the comparison to the accessibility of the lunar surface is a reasonable long-term goal. If we continue to use the sample of the known NEOs, we find ~1100 objects of all sizes with delta-v < 5.1 km/s (similar to the delta-v needed to reach (101955) Bennu in the Benner list), only 78 of which have H < 22.5. This suggests ~5-6 Ch-type objects larger than ~100 m diameter should be easier to reach than Bennu and already discovered (if not known to be Ch-class). Whether this number is too low to support an asteroid mining economy, or to get one started, is a question left for venture capitalists and angel investors rather than one for planetary scientists.

## 7.2. NASA Ames Trajectory Browser-based estimates

The NASA Ames Trajectory Browser (hereafter just "Trajectory Browser": https://trajbrowser.arc.nasa.gov/index.php) allows users to search for trajectories to small bodies meeting desired criteria for object size, total delta-v, mission duration, etc. It provides more rigorous results than discussed in Section 7.1, but it also suffers from some drawbacks for our purposes: It only returns 100 trajectories even if more than 100 fit the criteria, and it does not report how many compliant trajectories exist. It also restricts the maximum mission duration to 10 years for a round-trip rendezvous and does not search for launch windows outside of the 2010-2040 timeframe. The maximum number of trajectories returned is the most serious limitation for this work, and requires us to extrapolate to reach some of the comparisons we wish to make.

We look first at the most accessible objects. The Trajectory Browser returns 58 objects of H < 22.5 with delta-v of 5.3 km/s or lower, comparable to (and including) Bennu as calculated by the Trajectory Browser. This implies 2-5 Ch asteroids (±1 σ), slightly fewer than the value from the Shoemaker-Helin calculations but within uncertainties.

As with the objects in the previous section, the results from the Trajectory Browser assume a direct return to Earth rather than a return to a high Earth orbit. We again use 9 km/s as the delta-v limit we use for an apples-to-apples comparison. More than 100 trajectories exist for H < 17.5 and a round-trip delta-v of 9 km/s or less. Increasing the minimum object size to H < 16.5 (roughly 3.2 km for the albedo discussed above) leads to 92 objects meeting the delta-v and other criteria. We increased the size limit further, with 56 objects returned for H < 16.0 and 16 objects for H < 15.0. Given these numbers, we fit a power law: log N(delta-v<9) = 0.5118H-6.465. We then tested this power law at other H values, with good results: 5 objects are predicted for H < 14.0 and 4 are returned, 75 are predicted for H< 16.3 and 75 are returned.

The NEO size-frequency distribution follows a power law fairly closely between H=12 and H=19 (Harris and D'Abramo 2015), so we can extrapolate this to objects with H < 17.5 (again, diameter ~1 km) fairly confidently. The extrapolated number of 1-km objects with delta-v < 9 km/s is 310 (compare to 270 in the previous section), implying 19 ± 9 Ch asteroids larger than 1 km are more accessible than the lunar surface, within 20% of the number calculated in the previous section.

We are much less confident extrapolating the power law to H < 22.5. Models of the NEO population deviate strongly from power-law behavior at H > 19, with the deviation strongest near the sizes of interest. Power-law behavior overpredicts the Harris and D'Abramo model SFD by a factor of 3.7 at H=22.5. Before any corrections, the power law predicts 112,000 objects > 100 m with delta-v < 9 km/s. Reducing this by a factor of 3.7 to approximate the Harris and D'Abramo SFD results in ~30,000 objects within those size and delta-v bounds. The implied Ch fraction for the *entire* population is 1800 ± 900. Because we have only discovered ~20% of the objects > 100 m and subject to the same caveats discussed above about assuming a well-mixed population in composition and delta-v, this would imply that 360 ± 180 Ch asteroids are known (if as yet uncharacterized) and more accessible than the lunar surface. This is larger than the results from Section 7.1 but, again, well within a factor of 2.

We summarize the estimated numbers of Ch asteroids in the populations discussed above in Table 2.

## 8. How Much Water?

We might try to go one step further still. Can we determine the mass of water that should be present in the NEO population? To do so, we must piece together an additional set of estimates. Rivkin et al. (2015b) found the average H:Si ratio for Ch asteroids to be 1.56 ± 0.76, which corresponds to an equivalent average water concentration of ~7 wt % using the relations and assumptions found in Rivkin et al (2003). The same work found the

H:Si ratio of CM meteorites to span the same range as the Ch asteroids whether estimated spectroscopically or measured in the laboratory. Measurements of CM2 meteorites by Howard et al. (2013) found phyllosilicate abundances in a narrow range from 73-79%, corresponding to H:Si ratios of ~1.8-1.9. This is slightly higher than, but within the uncertainties of, what is estimated for the Ch asteroids. We adopt 7 wt % water for a typical Ch asteroid, with the recognition that this may be an underestimate and that the representativeness of asteroidal regolith measurements when studying asteroidal interiors is very much an open question.

Most of the hydrated mass of the NEOs will be found in the volume of the largest (or the few largest) hydrated NEOs. At present, these are (285263) 1998 QE2, (100085) 1992 UY4, and (175706) 1996 FG3, with diameters of 3.2 ± 0.3 (Springmann et al. 2014), 1.68 ± 0.08 (Volquardsen et al. 2007) and 1.71 ± 0.07 km (Wolters et al. 2011), respectively. Because 1992 UY4 is not a C-complex asteroid, we omit it for consistency with the rest of this discussion. The other two asteroids, interestingly both binary systems with ~500-m satellites, are also both more accessible than the surface of the Moon. If we assume both are spherical, the equivalent mass of water ($m_{H2O}$) contained in each object is easily seen to be:

$$m_{H2O} = (4/3)\pi(D/2)^3 \rho f_{H2O}$$

where $\rho$ is the density and $f_{H2O}$ is 0.07 from above. Both objects also have measured densities because they are binary systems: 700 ± 200 kg/m$^3$ for 1998 QE2 (Springmann et al. 2014) and 1300 ± 170 (1-$\sigma$) kg/m$^3$ (Scheirich et al. 2015) for 1996 FG3. The resulting mass of water is $8.4 \times 10^{11}$ kg for the primary of 1998 QE2 and $2.4 \times 10^{11}$ kg for the primary of 1996 FG3, with values ~100 times smaller for the satellites. These masses of water would have equivalent diameters of 450 and 296 m, respectively. The uncertainties in these numbers are ~30% for 1998 QE3 and ~5% for 1996 FG3.

A 1-km object with 7% water and a density like that of 1996 FG3 would have $4.8 \times 10^{10}$ kg of water. If we take the smaller estimates from the previous section, this would imply ~$8 \times 10^{11}$ kg (±50%) of water in the 1-km NEO population that is more accessible than the Moon.

## 9. Implications for Lunar Ice

We now briefly return to one of the scientific drivers for studying hydrated asteroids. It has been established that the lunar poles host permanently shadowed regions where ice is stable and found (Arnold 1979, Vasavada et al. 1999, Colaprete et al. 2010). It is generally thought that this ice was delivered to the Moon via impact, with a small fraction of water molecules making their way to the cold traps and the rest escaping or being destroyed (Crider and Vondrak, 2003). Estimates of the mass of ice at the lunar poles range from $1.98 \times 10^{11}$ kg (Feldman et al. 2000) to $7.4 \times 10^{11}$ kg (Colaprete et al. 2010).

Objects of 1 km diameter strike the Earth roughly every 500,000 years (Harris and D'Abramo 2015), and using the 6% Ch NEO fraction derived above we can expect 1-km Ch asteroids to impact roughly every 8 My. The Moon is struck ~60% less often than the

Earth (Le Feuvre and Wieczorek 2011), increasing that frequency to 13 My. As seen above, 1-km Ch asteroid contains ~5 × $10^{10}$ kg of water, and in principle ~15 such objects are needed to bring enough water to account for the water at the lunar poles for the Colaprete et al. mass estimate. However, the efficiency of water retention and transport in such a case is not obvious. Studies of water retention after cometary impacts suggests additional losses, with only 10% of water delivered retained on the Moon, only 5% of those molecules making it to lunar cold traps, and only ~5% of molecules in cold traps surviving long-term (Stewart et al. 2011, Ong et al. 2010, Crider and Vondrak 2003). These suggest that only 0.025% of water that is delivered will make it to the cold traps, increasing the number of necessary impacts to roughly 60,000 and the timescale for filling the cold traps to one much longer than the age of the Solar System. It is conceivable that the smaller impact speeds of asteroids compared to comets might lead to greater retention of water upon impact than 10%, but the subsequent loss mechanisms would remain as modeled, and would still disfavor accumulation of asteroidal water.

At the other extreme, there are reasons to expect asteroidally-delivered water to remain at its impact site. The Bench Crater CM meteorite, found in the Apollo 12 returned sample, retains intact hydrated minerals (Zolensky 1997), suggesting that water may remain in any asteroidal minerals delivered by impact rather than being transported further (and/or destroyed en route). This case is also inconsistent with hydrated asteroids being a major contributor to the lunar cold traps, though the equivalent to several cold traps worth of water may be spread across the lunar surface still in hydrated minerals.

## 10. Conclusions and Future Work

We have estimated the number of Ch-class asteroids in near-Earth space, of interest for several science and non-science reasons. Because Ch asteroids are thought to be analogous to CM meteorites, the most common hydrated meteorite type, we take the Ch fraction to be an estimate of the hydrated asteroid fraction. NEO delivery models suggest that 17% ± 3% of the NEOs should be Ch class, but estimates based on asteroids observed in the NEO population lead to a range that averages less than half that number: 6% ± 3%. This discrepancy is rooted in the difference between the C-complex population expected and observed in the NEO population. There are a variety of possible reasons for the discrepancy, ranging from biases in delivery from the main belt and in destruction of objects to observational bias. There are additional factors at work, likely atmospheric screening, that make the CM fraction of meteorite falls smaller still.

Using conservative estimates for the Ch fraction in near-Earth space, we calculate that hundreds of known NEOs should be hydrated and also more accessible than the surface of the Moon. We estimate that ~300 ± 150 of these objects are 100 m in diameter or larger. In a statistical sense, there should be many hydrated asteroids among the known NEOs, with identifying them the main missing piece of the puzzle. We expect measurements from Gaia to aid matters by better measuring or constraining the Ch fraction in the main asteroid belt (Delbo et al. 2012). Directly identifying Ch-class NEOs likely awaits a dedicated survey program of visible-wavelength measurements.

## 11. Acknowledgements

ASR was supported for this work by NASA grant NNX14AL60G from the Near-Earth Object Observations program and the VORTICES team of the Solar System Exploration Research Virtual Institute (NNA14AB02A). Reviews by Lucy McFadden and an anonymous referee improved this work significantly. Thanks to Brent Barbee and Chris Lewicki for useful discussions. ASR and FED served as science advisors for Planetary Resources Incorporated, though they received no compensation or support for this work. Part of the data utilized in this publication were obtained and made available by the MIT-UH-IRTF Joint Campaign for NEO Reconnaissance. The data in Figure 1 came from multiple sources, and can be recreated from those sources: 0.43-2.5 $\mu$m data is available for both 2 Pallas and 13 Egeria at the MIT-UH-IRTF Joint Campaign website (http://smass.mit.edu/catalog.php), 2-4 $\mu$m data for Pallas and Egeria can be found at the JHU-APL Data Archive (http://lib.jhuapl.edu/papers/asteroid-21-lutetia-at-3-micrometers-observations-/ for Pallas and http://lib.jhuapl.edu/papers/the-ch-class-asteroids-connecting-a-visible-taxono/ for Egeria, respectively).

**Tables**

| Parameter | Source | Value |
|---|---|---|
| Ch fraction of C complex for regions of main belt | R12, F14 average | 39 ± 10%, 42 ± 6 %, 32 ± 6% (inner, mid, outer) |
| Fraction of NEOs contributed from regions of main belt | B05 | 61 ± 9%, 24 ± 5%, 8 ± 1% (inner, mid, outer) |
| Fraction of all asteroids that are C complex for regions of main belt | B02, D16 average | 43 ± 5%, 50 ± 3%, 59 ± 7% (inner, mid, outer) |
| **Fraction of NEOs of given class/complex** | **R12/F14 + B05 + B02/D16** | **17 ± 3% Ch, 43 ± 6% C** |
| Ch fraction of C-complex NEOs | B04 | 4% (1 of 23) |
| C-complex fraction of NEOs | S04, C16 average | 16 ± 7% |
| **Estimated Fraction of Ch NEOs** | **R12/F14 + S04/C16** | **6 ± 3%** |
| CM fraction of meteorite falls | Met. Bull. | 1% |

**Table 1:** Input parameters used for estimating the fraction of hydrated (Ch-class) NEOs, and the resulting (bolded) estimates. Source citation codes are as follows: R12 = Rivkin (2012), F14 = Fornasier et al. (2014), B05 = Bottke et al. (2005), B02 = Bus and Binzel (2002), D16 = DeMeo and Carry (2016). B04 = Binzel et al. 2004, S04 = Stuart and Binzel (2004), C16 = Carry et al. (2016), Met Bull = Meteoritical Bulletin (https://www.lpi.usra.edu/meteor/metbull.php, 9 Feb 2018 update).

| Population | Total Number | Estimated number of Ch asteroids in population |
|---|---|---|
| NEOs > 1 km | 886 (known) | 53 ± 27 |
| NEOs > 1 km, Δv < Moon | 270 (S-H) <br> 310 (TB) | 16 ± 8 (S-H) <br> 19 ± 9 (TB) |
| NEOs > 100 m, Δv < Moon | 4487 (S-H) <br> ~6000 (TB) | 269 ± 135 (S-H) <br> ~360 ± 180 (TB) |
| All NEOs with Δv < Moon | 11,785 (S-H) | 700 ± 350 (S-H) |
| NEOs > 100 m, Δv ≤ Bennu | 78 (S-H) <br> 58 (TB) | ~5-6 (S-H) <br> ~2-5 (TB) |

**Table 2:** Summary of estimated number of Ch asteroids for different NEO sub-populations. The 100-m size range includes objects with H < 22.5, which represents 100 m sizes for average NEO albedos, but is > 100 m for C-complex albedos (see text for fuller discussion). The values estimated using delta-v generated via the Shoemaker and Helin approach are denoted "S-H", those using the NASA Ames Trajectory Browser with (TB). The column with total number of asteroids is based on discoveries and estimates as of 1 September 2018.